\documentstyle[psfig]{caosp}
\begin{document}

\title{Dust and Gas around $\lambda$ Bootis Stars}
\author{I. Kamp \and H. Holweger} 
\institute{Institut f\"{u}r Theoretische Physik und Astrophysik\\
           Universit\"{a}t Kiel\\Germany}

\date{January 29, 1998}
\maketitle

\begin{abstract}
High-resolution spectra of $\lambda$ Bootis stars reveal the
presence of circumstellar gas for example in the Ca K line.  

The example of the normal A star $\beta$ Pictoris shows, that the 
narrow stable absorption component in Ca K can be reproduced using 
appropriate disk models and a calcium underabundance in the circumstellar 
gas of a factor of $\sim 30$.

Similar models are suggested for the group of metal-deficient
$\lambda$ Bootis stars, but the observational material is still
very poor.
\end{abstract}

\section{Introduction}

High-resolution spectrometry and NLTE analysis of $\lambda$ Bootis stars
reveals the metal-deficient nature of this small subgroup of main-sequence
A stars (Venn \& Lambert 1990, Holweger \& St\"{u}renburg 1993,
St\"{u}renburg 1993). Venn \& Lambert (1990) suggested that the abundance
anomalies of these stars are due to accretion of circumstellar (CS) gas, 
which is depleted in condensable elements.

Evolutionary tracks for the $\lambda$ Bootis stars and a recent 
high-S/N search for circumstellar Ca K lines raise the question 
of a possible pre-main-sequence evolutionary status for these stars 
(Gerbaldi et al. 1993, Holweger \& Rentzsch-Holm 1995, Paunzen et al. 
1998). In this case the circumstellar gas may have remained 
from the star formation phase. 

The disk models presented in the following section are in close
analogy to those for more massive disks around T-Tauri stars.
A short discussion has been given by Rentzsch-Holm, Holweger
\& Bertoldi (1998).

\section{Circumstellar disk models}
\label{csmod}

The gas density distribution of our model corresponds to thin disks
in hydrostatic equilibrium:
\begin{equation}
\rho (r) = \rho_{\rm i} \cdot \left(r\over R_{\rm i}\right)^{-\epsilon} 
           ~e^{-(h / H r)^2} ~,
\end{equation}
where $R_{\rm i}$ is the inner disk radius, $r$ the radial distance from
the star, $h$ the vertical disk height above the disk, and
$H\equiv \sqrt{2} h_0/r\approx$ constant is the scale height scaled to the radius.
The outer disk radius, $R_{\rm o}$, and total disk mass then determine the midplane
density at the inner disk radius, $\rho_{\rm i}$. The dust responsible
for UV absorption is assumed well mixed by turbulence, and 
$H_{\rm gas} = H_{\rm dust} \sim 0.2$ (Pringle 1981). 

The disk temperature is derived from radiative equilibrium calculations
using dust opacities from Henning \& Stognienko (1996) for a mixture
of various grain materials and sizes of up to 5~$\mu$m.

\begin{figure}[hbt]
\psfig{figure=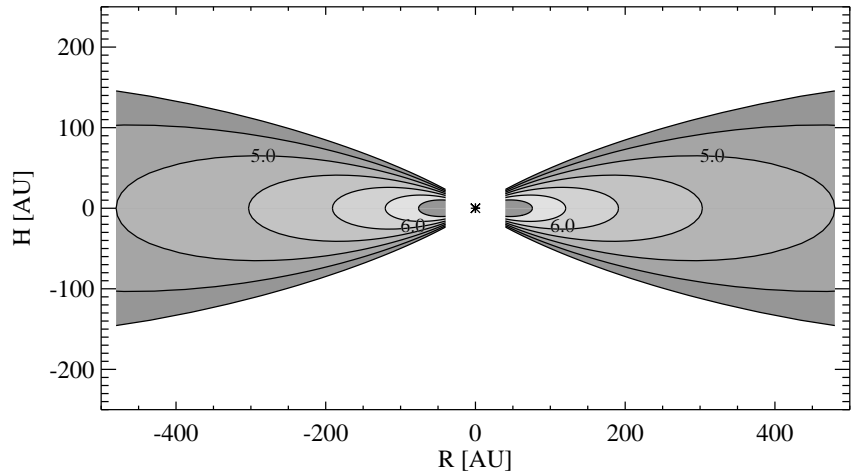,height=7cm}
\vspace{2mm}
\caption{Density distribution in the $\beta$ Pictoris model: 
 $R_{\rm i}=40$~AU, $R_{\rm o}=480$~AU, $M_{\rm tot}=46$~M$_{\rm E}$, 
 \mbox{ATLAS9}(8000, 4.0, 0.0), $R_{\ast}=1.7 R_{\odot}$. Shown are
 isodensity lines on a logarithmic scale. The outer boundary is
 defined by the disk density dropping to typical values of 
 molecular clouds, $10^4$~cm$^{-3}$ }
\label{f1}
\end{figure}

Figure~\ref{f1} shows the density distribution in a $\beta$ Pictoris
model. The disk extends from 40 to 480~AU and comprises a total mass of
46 earth masses ($M_{\rm gas}:M_{\rm dust}=100$). The parameters are chosen 
in accordance with the results of Chini et al. (1991). 

\section{The circumstellar Ca\,{\sc ii}-K lines}

Holweger \& Rentzsch-Holm (1995) found narrow absorption components in 5
of the 11 $\lambda$ Bootis stars studied. This suggests that the 
$\lambda$ Bootis phenomenon is related to the presence of circumstellar 
gas. Because of their faintness $\lambda$ Bootis stars were hardly
detectable by IRAS and hence information on the presence and extension of 
circumstellar dust and its mass is very poor. 

In the following the prominent A star $\beta$ Pictoris is taken as an 
example for the formation of a narrow absorption component in a  
circumstellar disk even though $\beta$ Pictoris itself does not 
belong to the group of $\lambda$ Bootis stars (Holweger et al. 1997).

The ionization equilibrium of Ca in the disk is determined by photoionization 
due to the unshielded stellar radiation field (Verner et al. 1996)
and recombination via collisions with electrons (Spitzer 1978).
Line broadening is due to thermal velocities.

\begin{figure}[hbt]
\psfig{figure=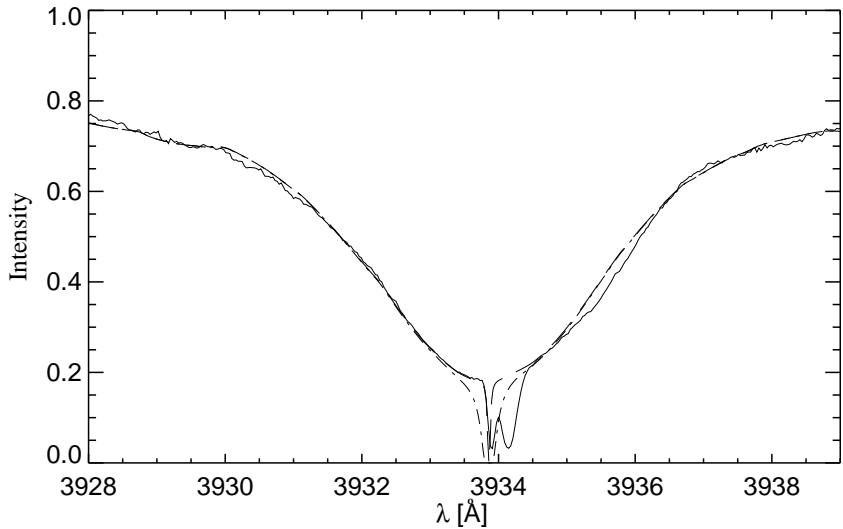,height=8cm}
\vspace{2mm}
\caption{Core of the Ca K line in $\beta$ Pictoris. Solid line:
         observation 1993 JAN 25, see Holweger \& Rentzsch-Holm (1995); 
         dashed line: photospheric calcium abundance [Ca/H]$=0.02$,
         circumstellar calcium abundance [Ca/H]$=-1.5$; dashed-dotted line: 
         photospheric calcium abundance [Ca/H]$=0.02$, circumstellar 
         calcium abundance solar }
\label{f2}
\end{figure}
Fig.~\ref{f2} shows the $\beta$ Pictoris Ca K line recorded
in 1993 with the Coud\'{e} Echelle Spectrograph of the ESO 1.4-m CAT
telescope at a resolving power of $50\,000$. The dashed line
represents a synthetic spectrum calculated with an 
\mbox{ATLAS9}(8200, 4.24, 0.0)
model atmosphere, a calcium abundance of [Ca/H]=0.02 and a $v \sin i$
of 132~km~s$^{-1}$ (Holweger \& Rentzsch-Holm 1995). The narrow
absorption component at the bottom of the rotationally broadened photospheric
line results from absorption by ionized calcium in the circumstellar 
disk model described in Sect.~\ref{csmod}. The disk is assumed to be
seen edge-on and calcium strongly depleted in the disk, [Ca/H]$=-1.5$. The 
dashed-dotted line represents the same model with a solar calcium
abundance in the disk.

\begin{figure}[hbt]
\psfig{figure=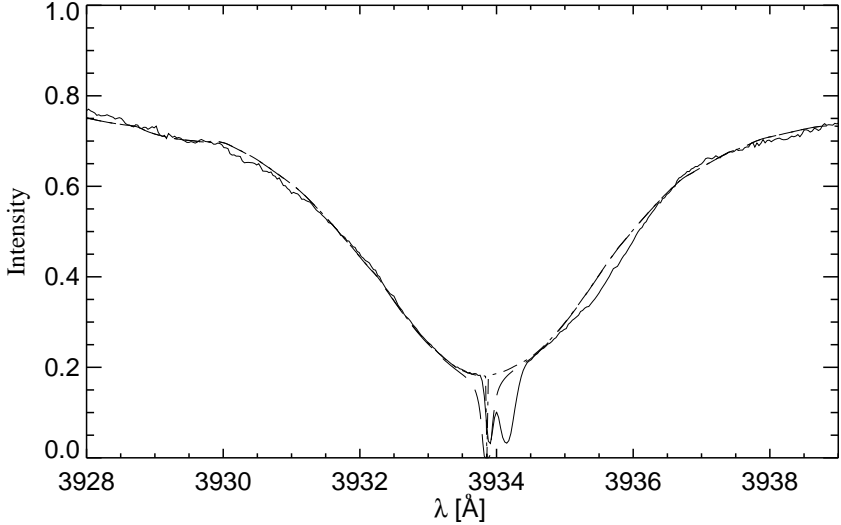,height=8cm}
\vspace{2mm}
\caption{Core of the Ca K line in $\beta$ Pictoris. Solid line:
         observation 1993 JAN 25, see Holweger \& Rentzsch-Holm (1995); 
         dashed line: solar calcium abundance in the disk (like dashed-dotted 
         line in Fig.~\ref{f2}) but $\beta=10^{\circ}$; dashed-dotted line: 
         solar calcium abundance in the disk but $\beta=30^{\circ}$ }
\label{f3}
\end{figure}
Fig.~\ref{f3} shows the variation of the circumstellar line profile
with increasing aspect angle $\beta$ ($\beta=0$ corresponds to "edge-on"). 
Observation at larger $\beta$
results in a narrower CS line profile, because the density 
strongly decreases with increasing distance from the equatorial plane.
Hence this has the same effect as decreasing the calcium abundance. 
The CS component will disappear if the line of sight to the star
does not intersect the disk as it may be the case for Vega, which
seems to be a fast rotating $\lambda$ Bootis star seen pole-on
(Gulliver, Hill \& Adelman 1994, Holweger \& Rentzsch-Holm 1995).

\section{Conclusion}

For $\beta$ Pictoris the stable component of the CS absorption line
in Ca K can be explained by absorption in a gaseous disk in Keplerian
rotation around the star. Ca\,{\sc ii} is the dominant ionization stage
as already pointed out by Lagrange et al. (1995). Since the inclination of the 
disk is only a few degrees (Smith \& Terrile 1984), the results suggest that 
calcium is underabundant by a factor of $\sim 30$ in the circumstellar 
disk around $\beta$ Pictoris.

The model presents the possibility to quantitatively study the
circumstellar lines observed in several $\lambda$ Bootis stars.
Nevertheless further information such as ISO observations or 
radio observations are needed to constrain the free 
parameters of the model, like disk masses and extensions. 

\acknowledgements

This work was supported by the "Deutsche Forschungsgemeinschaft"
under grant Ho 596/35-1.

\end{document}